\documentclass{article}
\usepackage{spconf,amsmath,graphicx}
\usepackage{pgfplots}
\usepackage{diagbox}
\usepackage{caption}
\usepackage{booktabs}
\usepackage{filecontents}
\usepackage{url}
\DeclareSymbolFont{toneletters}{T1}{\familydefault}{m}{it}
\SetSymbolFont{toneletters}{bold}{T1}{\familydefault}{bx}{it}

\DeclareMathSymbol\eth{\mathord}{toneletters}{"F0}

\vspace{5pt}
\title{Masked Autoencoders Are Articulatory Learners}
\name{Ahmed Adel Attia and Carol Y. Espy-Wilson\thanks{This work was supported by the National Science Foundation grant IIS1764010.}}
\address{University Of Maryland, College Park\\
Institute For Systems Research\\
Maryland, USA}
\begin{document}
\urlstyle{rm}

%\ninept
%
\maketitle
\begin{abstract}
Articulatory recordings track the positions and motion of different articulators along the vocal tract and are widely used to study speech production and to develop speech technologies such as articulatory based speech synthesizers and speech inversion systems. The University of Wisconsin X-Ray Microbeam (XRMB)  dataset is one of various datasets that provide articulatory recordings synced with audio recordings. The XRMB articulatory recordings employ pellets placed on a number of articulators which can be tracked by the microbeam. However, a significant portion of the articulatory recordings are mistracked, and have been so far unusable. In this work, we present a deep learning based approach using Masked Autoencoders to accurately reconstruct the mistracked articulatory recordings for 41 out of 47 speakers of the XRMB dataset. Our model is able to reconstruct articulatory trajectories that closely match ground truth, even when three out of eight articulators are mistracked, and retrieve 3.28 out of 3.4 hours of previously unusable recordings. 
\end{abstract}
\vspace{-7pt}
\section{Introduction}
\label{sec:intro}
Articulatory data have been foundational in helping speech scientists study the articulatory habits of speakers that can differ between individuals, and across languages and dialects of the same language.  At present, specialized equipment such as X-ray Microbeam (XRMB) \cite{xrmb} Electromagnetic Articulometry (EMA) \cite{ema} and real-time Magnetic Resonance Imaging (rt-MRI) \cite{mri} is needed to observe articulatory movements directly to help researchers and clinicians understand these habits. Regardless of the technique, current technologies used in articulatory recordings sometime fail to correctly capture one or more of the articulators, resulting in mistracked segments.  Given the considerable time and cost involved in data collection and the potential prolonged exposure to harmful signals, these mistracked pellets are seldom re-recorded leaving some recordings unusable.
\vspace*{-1.9pt}

There has been considerable work to develop techniques to reconstruct missing articulatory recordings. In \cite{density-xrmb}, the authors proposed an algorithm for recovering missing data by learning a density model of the vocal tract shapes. Their model was limited to recovering only one mistracked articulator at a time. Also, their approach requires handcrafting the model parameters for every speaker and every articulator. In \cite{gosh_kalman, gosh-MAE}, a Kalman smoother and maximum a-posteriori estimator fills in the missing samples. Their work was also limited to one articulator at a time, and was additionally dependent on the length of the mistracked samples. 

Our approach does not suffer from the same limitations of previous works. The model hyper-parameters do not require any handcrafting for different speakers, and one model can predict any missing articulator. As a result, our model is easily applicable for the majority of the dataset. We are also not limited to reconstructing one missing articulator, or limited to a certain duration of mistracked segments, but are able to reconstruct up to three missing articulators that follow the ground truth closely for the entire duration of the recording. 

We approach the problem as a masked reconstruction problem. Recently, there has been some major breakthroughs in image reconstruction using masked autoencoders \cite{mae}, where an image can be realistically reconstructed with 90\% of it being masked in patches. This work follows a similar approach with articulatory data, where a portion of an input frame is masked and then fed to an autoencoder, which then reconstructs the entire frame including the masked portions. We explain our approach in greater detail in section \ref{sec:approach}. Section \ref{sec:data} outlines the dataset description. We demonstrate test set results in section \ref{sec:results}, and discuss the limitations of the proposed approach in \ref{sec:limitations}. We end with a conclusion and discussion about future directions in section \ref{sec:conclusion}.
\vspace{-10pt}
\section{Dataset description}
\vspace{-10pt}
\label{sec:data}
In this study, we use the University of Wisconsin XRMB dataset. However, there is no reason the proposed approach cannot be applied to other datasets. 
\subsection{XRMB dataset}
The XRMB dataset contains articulatory recordings synced with audio recordings. Each speaker had 8 gold pellets glued on 8 articulators, upper lip (UL), lower lip (LL), tongue tip (T1), tongue blade (T2), tongue dorsum (T3), tongue root (T4), mandible incisor (MNI), and (parasagittally placed) mandible molar (MNM). The X-Y coordinates for every pellet were tracked through time and sampled at different sampling rates. All pellet trajectories (PTs) were then resampled to have a sampling rate of 160 samples per second.

We only use data from 41 speakers from the XRMB dataset. These speakers have enough uncorrupted data for the model to train and generalize. The six speakers not used (JW29, JW32, JW33, JW42, JW44, JW60) do not have enough uncorrupted data for training.

All recordings start with a tone to show the speaker when to start speaking, and some recordings contain trailing silences at the end of the recording, and mid-sentence silences between different utterances. We trim the silences from these utterances. Some tasks do not include speech (swallowing, ...etc). We do not include nonverbal tasks in the dataset.

\begin{figure}[]
    \centering
\resizebox{0.7\columnwidth}{!}{%
\begin{tikzpicture}
\begin{axis}[
    ybar stacked,
	bar width=50pt,
    ymin =0,
    legend style={at={(0.5,0)},
      anchor=north,legend columns=-1},
    ylabel={Percentage},
    symbolic x coords={0},
    xtick= \empty
    ]
\addplot+[ybar] plot coordinates {(0,72.78)};
\addplot+[ybar] plot coordinates {(0,20.35)};
\addplot+[ybar] plot coordinates {(0,5.06)};
\addplot+[ybar] plot coordinates {(0,1.81)};
\legend{\strut One, \strut Two, \strut Three, \strut More Than Three}
\end{axis}
\end{tikzpicture}
}
\caption{Percentage of total duration of recordings with different degrees of mistracking to the total duration of recordings with mistracked samples in the XRMB dataset.}
\vspace*{-20pt}
\label{fig:bar_stacked}
\end{figure}
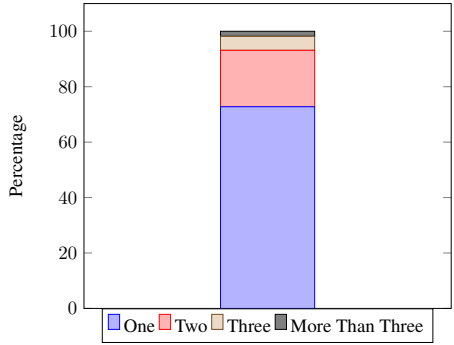

Table \ref{table: dataset_calc} shows the amount of mistracking in different subsets of the XRMB dataset. The third column shows the duration of data if we limit ourselves to only using recordings without any mistracking. Such a limitation is reasonable to ensure that the recordings can be broken down into contiguous frames. Most discarded recordings contain only a few milliseconds of mistracking, but are rendered useless. Figure \ref{fig:bar_stacked} shows that more than 25\% of the files with mistracking present have multiple PTs concurrently mistracked.

\vspace*{-10pt}
\begin{table}[]
  \caption{Percentage and duration of clean and mistracked data in different subsets of the XRMB dataset.}
  \vspace*{-5pt}
  \centering
  \resizebox{\columnwidth}{!}{\begin{tabular}{|c |c| c| c| c|}
    \hline
   \rule{0pt}{6ex} \textbf{\large{Dataset Subset}} & 
                    \textbf{\large{Total Duration Of Recordings}}
                                & \multicolumn{1}{c|}{\begin{tabular}{@{}c@{}}\textbf{\large{Total Duration Of Recordings}} \\ \textbf{\large{Without Mistracking}}\end{tabular}}
                                & \multicolumn{1}{c|}
                                {\begin{tabular}{@{}c@{}}\textbf{\large{\% Recordings With}} \\ \textbf{\large{ Mistracking}}\end{tabular}}
    \\[15pt]\hline
     \rule{0pt}{3ex} \textbf{\large{47 Speakers}} & {\Large 14.36 hrs} &  {\Large 8.62 hrs} &  {\Large 37.4}\% \\[5pt]
    \hline
     \rule{0pt}{5ex} \textbf{\begin{tabular}{@{}c@{}}\textbf{\large{41 Speakers used}} \\ \textbf{\large{in this study}}\end{tabular}} & {\Large 12.68 hrs} &  {\Large 8.46 hrs} &  {\Large 30.3\%} \\[10pt]
    \hline
     \rule{0pt}{6ex} \begin{tabular}{@{}c@{}}\textbf{\large{41 Speakers used in this study}} \\ \textbf{\large{using only the verbal tasks}} \\ \textbf{\large{with extended silences removed}}\end{tabular} & {\Large 10.6 hrs}  & {\Large 7.2 hrs}&  {\Large 30.3\%}
    \label{table: dataset_calc}
    \\[16pt]\hline
  \end{tabular}}
% \vspace*{-10pt}
\end{table}

\subsection{Training and test sets}
The training data was created through a sliding window on all recordings. After the frames were created, only the frames without any mistracking were kept and the rest were disregarded. The test set was created similarly. Instead of randomly sampling the test set from the entire dataset, a set task list was hand-crafted, where all the recordings corresponding to these tasks are used in the test set. This was done to ensure that the test set covers the span of all verbal tasks in the dataset, is a good representation of the out-of-sample performance and is as consistent as possible across speakers. Since not all speakers performed all tasks, a similar task was chosen instead when a speaker didn't perform a task in the predetermined task list. Separate training and test datasets were created for every speaker, and a different model was trained for each speaker.
% \subsection{Data augmentation}
% A simple data augmentation technique was applied to the training set to increase its perceived size. Each file in the training set was upsampled and downsampled by a random sampling rate uniformly chosen between 1 to 1.5 times the original sampling rate for upsampling, and between 0.5 and 1 times the original sampling rate for downsampling. Emperical experiments show that this data augmentation technique improve generalization and test set performance.
\vspace*{-13pt}
\section{Approach}
\label{sec:approach}
As mentioned in Section \ref{sec:intro}, we follow a similar approach to the work put forth in \cite{mae}, where masked autoencoders reconstructed images with large amounts of masking. Inspired by this work, we designed our deep learning model using time-dilated dense layers and Bi-directional Gated Recurrent Network layers (Bi-GRUs).

We use a 200 sample input frame. The input frames are masked randomly. For each batch, a masking layer samples an integer between 1 and 8 uniformly $N$ times, with $N$ being a hyper-parameter. Unique samples represent the index of PTs to be masked and replaced by a trainable mask token. This means that the hyper-parameter $N$ controls the maximum amount of masking during training.

Unlike \cite{mae}, we don't remove the masked patches from the input to the encoder and provide positional information to the input of the decoder. Instead, we mask in place. This ensures that separate channels in the model keep the information about the identity of different PTs, masked or unmasked.
The model was implemented with the TensorFlow-Keras machine learning framework \footnote{\url{https://github.com/ahmedadelattia/MAE_Articulatory_Learners}} and trained on an NVIDIA Titan Xp GPU. We used Adam optimizer and mean absolute error loss. We employed early stopping with a patience of 50 epochs (the training is terminated when the test set loss isn't improved for 50 epochs). Even though the aim of the model is to reconstruct masked PTs, the loss is calculated on the entire output tensor to ensure that the model keeps the information about  inter-dependencies between different PTs throughout. We trained 8 models for every speaker, varying $N$ from 1 to 8. In the end, we choose only one model for every speaker for inference based on test set performance. 
\vspace{-10pt}
\section{Results}
\label{sec:results}
\vspace{-7pt}
The results in this section approximate worst-case conditions, where one or more PTs is mistracked for the entire test set. The Pearson Product Moment Correlation (PPMC) score between the predicted and ground truth PTs is calculated and presented. In practical settings, PTs are commonly only mistracked for a short duration (50-500 ms)\cite{xrmb}. However, recordings where one PT is entirely mistracked for the full duration of the recordings exist, and hence we use this as our test case. Note that fixing short segments of mistracking will make entire recordings, each a few seconds long usable again.
\vspace{-10pt}
\subsection{Tuning hyper-parameter $N$}
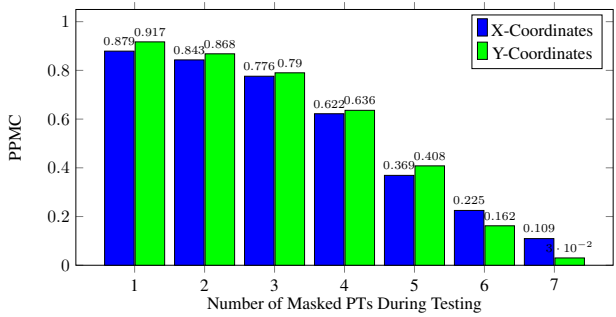
\begin{figure}[]
    \centering
\resizebox{0.95\columnwidth}{!}{%
    \begin{tikzpicture}[
      declare function={
        barW=4pt; % width of bars
        barShift=barW/2; % bar shift
      }
    ]
    \pgfplotsset{%
    height=0.38\textwidth,
    width=.7\textwidth
    }
    \begin{axis}[
        ybar = 0.6pt,
        ylabel=PPMC,
        xlabel = Number of Masked PTs During Testing,
        ymin =0, ymax=1.05,
        enlarge x limits = 0.14,
        bar width=17pt,
        symbolic x coords={1,2,3,4,5,6,7},nodes near coords,
        nodes near coords style={font=\scriptsize, /pgf/number format/.cd,precision=3},        
        xtick=data,         
        ytick={0,0.2,0.4,0.6,0.8,1},
        legend image code/.code={
        \draw [#1] (0cm,-0.1cm) rectangle (0.2cm,0.25cm); }]
    
        \addplot[ybar,fill=blue] coordinates {
            (1, 0.879)
            (2, 0.843)
            (3, 0.776)
            (4, 0.622)
            (5, 0.369)
            (6, 0.225)
            (7, 0.1094)
        };
        \addplot[ybar,fill=green] coordinates {
            (1, 0.917)
            (2, 0.868)
            (3, 0.790)
            (4, 0.636)
            (5, 0.408)
            (6, 0.162)
            (7, 0.03)
    };
    \legend{X-Coordinates, Y-Coordinates}
    \end{axis}
    \end{tikzpicture}
}
    \vspace*{-8pt}
    \caption{Average PPMC scores for different levels of masking during testing for a model trained with $N$ = 1.}
    \vspace*{-10pt}
    \label{barchart: N=1}
\end{figure}

\begin{figure}[]
    \centering
\resizebox{0.95\columnwidth}{!}{%
    \begin{tikzpicture}[
      declare function={
        barW=4pt; % width of bars
        barShift=barW/2; % bar shift
      }
    ]
    \pgfplotsset{%
    height=0.38\textwidth,
    width=.7\textwidth
    }
    \begin{axis}[
        ybar = 0.6pt,
        ylabel=PPMC,
        xlabel = Number of Masked PTs During Testing,
        ymin =0, ymax=1.05,
        enlarge x limits = 0.14,
        bar width=17pt,
        symbolic x coords={1,2,3,4,5,6,7},nodes near coords,
        nodes near coords style={font=\scriptsize, /pgf/number format/.cd,precision=3},        
        xtick=data,         
        ytick={0,0.2,0.4,0.6,0.8,1},
        legend image code/.code={
        \draw [#1] (0cm,-0.1cm) rectangle (0.2cm,0.25cm); }]
    
        \addplot[ybar,fill=blue] coordinates {
            (1, 0.847)
            (2, 0.836)
            (3, 0.819)
            (4, 0.786)
            (5, 0.724)
            (6, 0.617)
            (7, 0.416)
        };
        \addplot[ybar,fill=green] coordinates {
            (1, 0.904)
            (2, 0.894)
            (3, 0.876)
            (4, 0.847)
            (5, 0.802)
            (6, 0.717)
            (7, 0.530)
    };
    \legend{X-Coordinates, Y-Coordinates}
    \end{axis}
    \end{tikzpicture}
}
    \vspace*{-8pt}
    \caption{Average PPMC scores for different levels of masking during testing for a model trained with $N$ = 5.}
    \vspace*{-10pt}
    \label{barchart: N=5}
\end{figure}
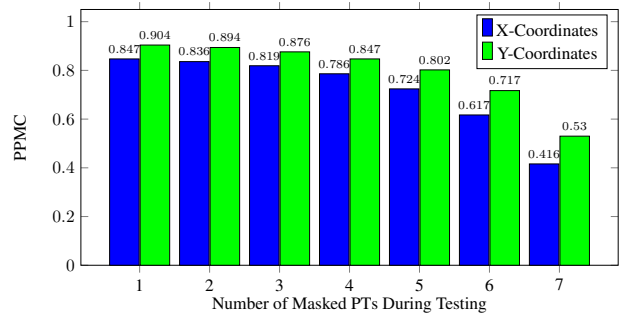

Empirical experiments showed that the higher the value of $N$, the better and more uniformly the model performs across different degrees of masking during testing. This means that by increasing $N$ the model performs better with higher masking levels during testing, even beyond the training upper-limit. However, values of $N$ beyond 4 usually lead to performance degradation for lower levels of masking during testing. Due to page limitations, we will not show comparisons between all 8 possible values of $N$, and we'll just demonstrate the difference between three different values for a single speaker (JW11) in figures \ref{barchart: N=1}, \ref{barchart: N=3} and \ref{barchart: N=5}.

\begin{figure}[]
    \centering
\resizebox{0.95\columnwidth}{!}{%
    \begin{tikzpicture}[
      declare function={
        barW=4pt; % width of bars
        barShift=barW/2; % bar shift
      }
    ]
    \pgfplotsset{%
    height=0.38\textwidth,
    width=.7\textwidth
    }
    \begin{axis}[
        ybar = 0.6pt,
        ylabel=PPMC,
        xlabel = Number of Masked PTs During Testing,
        ymin =0, ymax=1.05,
        enlarge x limits = 0.14,
        bar width=17pt,
        symbolic x coords={1,2,3,4,5,6,7},nodes near coords,
        nodes near coords style={font=\scriptsize, /pgf/number format/.cd,precision=3},        
        xtick=data,         
        ytick={0,0.2,0.4,0.6,0.8,1},
        legend image code/.code={
        \draw [#1] (0cm,-0.1cm) rectangle (0.2cm,0.25cm); }]
    
        \addplot[ybar,fill=blue] coordinates {
            (1, 0.883)
            (2, 0.870)
            (3, 0.850)
            (4, 0.808)
            (5, 0.726)
            (6, 0.588)
            (7, 0.376)
        };
        \addplot[ybar,fill=green] coordinates {
            (1, 0.928)
            (2, 0.915)
            (3, 0.894)
            (4, 0.858)
            (5, 0.791)
            (6, 0.672)
            (7, 0.463)
    };
    \legend{X-Coordinates, Y-Coordinates}
    \end{axis}
    \end{tikzpicture}
}
    \vspace*{-8pt}
    \caption{Average PPMC scores for different levels of masking during testing for a model trained with $N$ = 3.}
    \vspace*{-10pt}
    \label{barchart: N=3}
\end{figure}
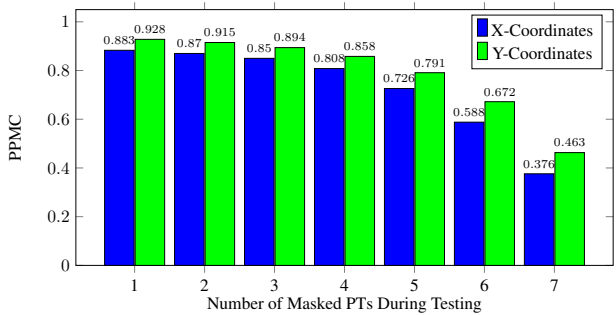

It should be noted that the overwhelming majority of corrupted recordings have at most 3 concurrently mistracked PTs, and hence the best values for $N$ will be chosen empirically for each speaker depending on the performance of each model when masking 1, 2 or 3 PTs concurrently.
\subsection{Overlapping frames for the training set}
We experimented with modifying the training data by allowing for 50\% overlap between adjacent frames. Overlap between frames increases the total number of frames, but in this case, the number of frames in the training data increased beyond the theoretical limit. This is due to the fact that we are utilizing recordings with mistracked samples by framing them and disregarding any frames with corruption, and keeping the rest. Overlapping increases the chance for previously unused samples to be caught in a sliding frame. Figure \ref{barchart: overlap, N=3} shows test set results for a model trained with overlapping frames and with $N$=3. Comparing with Figure \ref{barchart: N=3} which shows results for an identical model trained on the same training set without overlap, we can see that allowing for overlap between adjacent frames in the training set increased test set performance by approximately 10 points on average. 

\begin{figure}[]
    \centering
\resizebox{0.95\columnwidth}{!}{%
    \begin{tikzpicture}[
      declare function={
        barW=4pt; % width of bars
        barShift=barW/2; % bar shift
      }
    ]
    \pgfplotsset{%
    height=0.38\textwidth,
    width=.7\textwidth
    }
    \begin{axis}[
        ybar = 0.6pt,
        ylabel=PPMC,
        xlabel = Number of Masked PTs During Testing,
        ymin =0, ymax=1.1,
        enlarge x limits = 0.14,
        bar width=17pt,
        symbolic x coords={1,2,3,4,5,6,7},nodes near coords,
        nodes near coords style={font=\scriptsize, /pgf/number format/.cd,precision=3},        
        xtick=data,         
        ytick={0,0.2,0.4,0.6,0.8,1},
        legend image code/.code={
        \draw [#1] (0cm,-0.1cm) rectangle (0.2cm,0.25cm); }]
    
        \addplot[ybar,fill=blue] coordinates {
            (1, 0.960)
            (2, 0.954)
            (3, 0.946)
            (4, 0.931)
            (5, 0.905)
            (6, 0.710)
            (7, 0.172)
        };
        \addplot[ybar,fill=green] coordinates {
            (1, 0.975)
            (2, 0.969)
            (3, 0.958)
            (4, 0.936)
            (5, 0.897)
            (6, 0.716)
            (7, 0.211)
    };
    \legend{X-Coordinates, Y-Coordinates}
    \end{axis}
    \end{tikzpicture}
}
    \vspace*{-8pt}
    \caption{Average PPMC scores for different levels of masking during testing for a model trained with $N$ = 3 and overlapping frames.}
    \vspace*{-10pt}
    \label{barchart: overlap, N=3}
\end{figure}
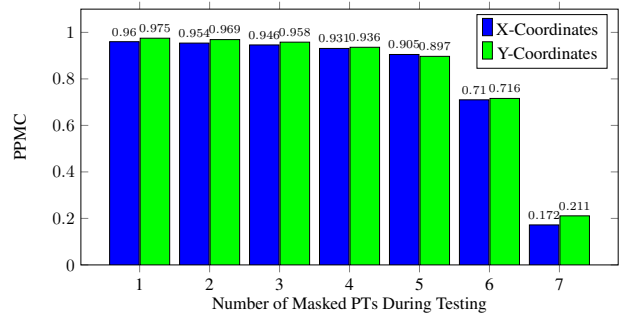
\subsection{Sample outputs}
\begin{figure}[t]
    \centering
    \includegraphics[width = \columnwidth]{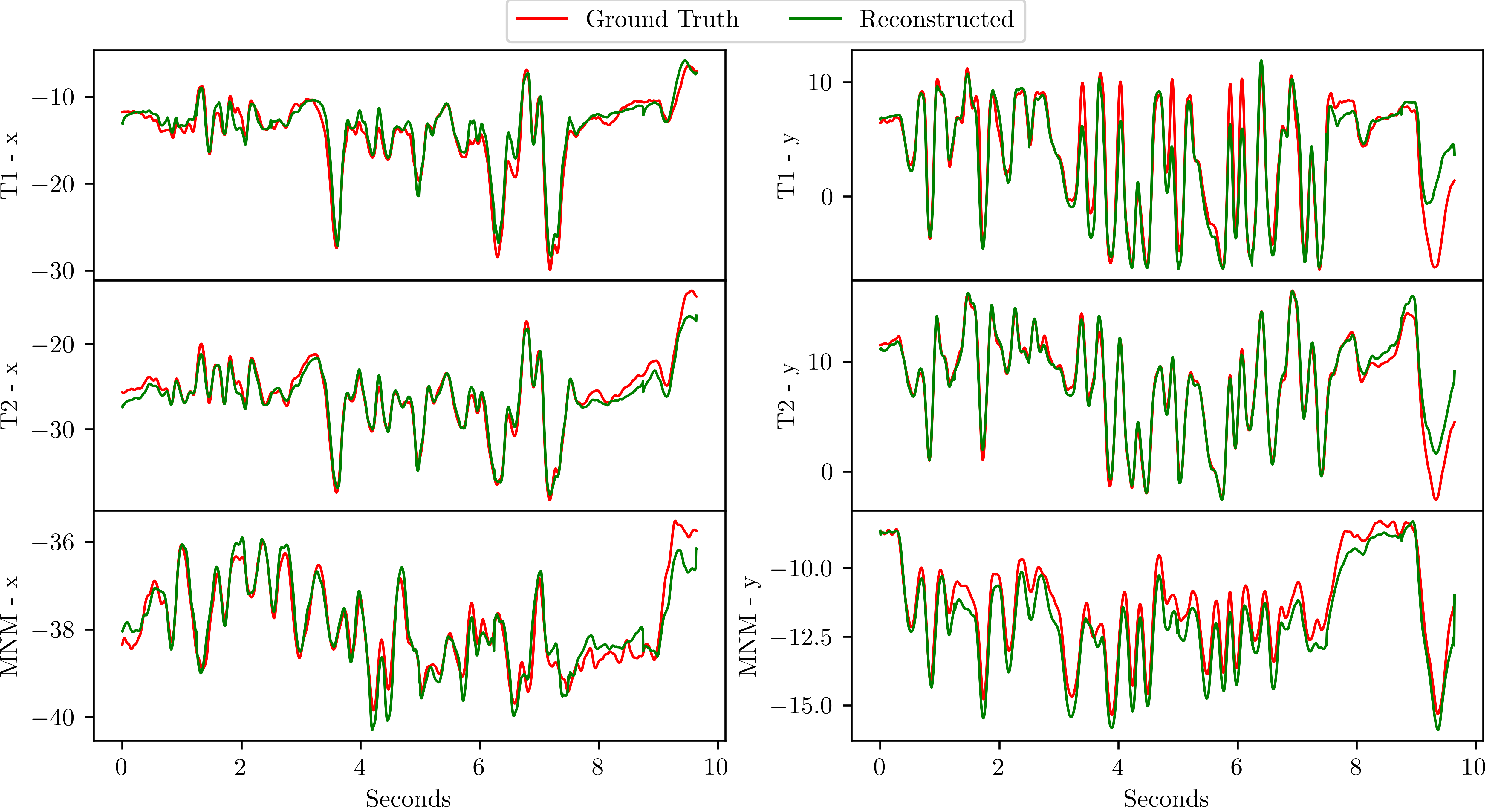}
    \vspace*{-8mm}
    \caption{Sample reconstruction of  \textbf{3 artificially and concurrently masked} PTs from task 003 (reading a paragraph) by speaker JW52.}
    \label{fig:sample}
\end{figure}

Figure \ref{fig:sample} shows a sample reconstruction for artificially masked data. Three PTs were randomly chosen and artificially masked for the entire duration of the recording (10 seconds). There is no instance in the XRMB dataset where three PTs are concurrently mistracked for such a long duration, so this test case provides less context to the model than actual use cases and is thus harder. The model was still able to closely follow the ground truth values of the PTs. 

\begin{figure}[t]
    \centering
    \includegraphics[width = \columnwidth]{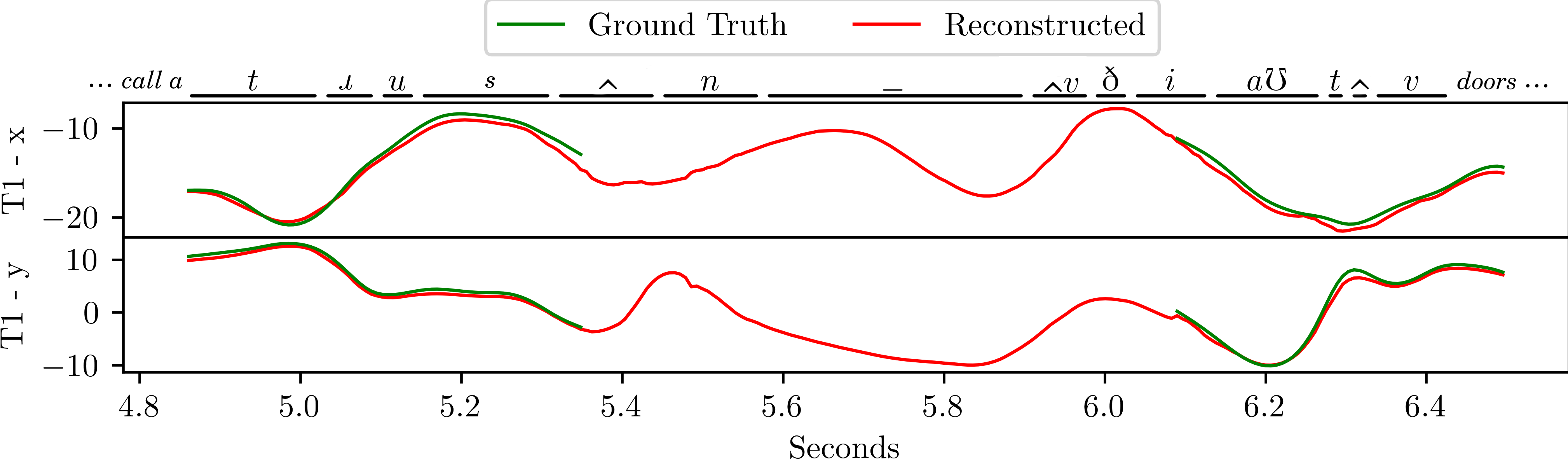}
    \vspace*{-8mm}
    \caption{Sample reconstruction of \textbf{actual corrupted} PTs from the utterance \textit{"...true son of the out of..."} from task 78 by speaker JW11.}
    \vspace*{-4mm}
    
    \label{fig:sample_corrupted}
\end{figure}

Figure \ref{fig:sample_corrupted} shows a sample reconstruction for actual corrupted data. The reconstructed trajectory of the mistracked segment matches the phonetic transcription, for instance, the peak in T1-y after 5.4 seconds corresponds to the $n$ sound, and the peak in T1-x at 6.0 seconds corresponds to the $\eth$ sound. Slight discontinuities can be seen between 5.4 and 5.6 seconds and 6.0 and 6.2 seconds at the edges of different frames and can be mitigated with a smoothing filter.
\vspace{-13pt}

\section{Performance Limitations}
\label{sec:limitations}
\begin{figure}[]
    \centering
\resizebox{0.95\columnwidth}{!}{%
    \begin{tikzpicture}[
      declare function={
        barW=4pt; % width of bars
        barShift=barW/2; % bar shift
      }
    ]
    \pgfplotsset{%
    height=0.38\textwidth,
    width=.7\textwidth
    }
    \begin{axis}[
        ybar = 0.6pt,
        ylabel=PPMC,
        xlabel = Masked PTs,
        ymin =0, ymax=1.2,
        enlarge x limits = 0.14,
        bar width=15pt,
        legend style={at={(0.51,1.2)},
        anchor=north,legend columns=0},
        symbolic x coords={UL, LL, T1, T2, T3, T4, MNI, MNM},
        nodes near coords,
        nodes near coords,style={font=\scriptsize, /pgf/number format/.cd,precision=2},   
        every node near coord/.append style={yshift = 0.25cm},
        xtick=data,         
        ytick={0,0.2,0.4,0.6,0.8,1},
        legend image code/.code={
        \draw [#1] (0cm,-0.1cm) rectangle (0.2cm,0.25cm); }]
    
        \addplot[ybar,fill=blue] plot [error bars/.cd, y dir= both, y explicit]
        table [y error plus = Corr-Max, y error minus = Corr-Min]{\xtable};
        \addplot[ybar,fill=green] plot [error bars/.cd, y dir= both, y explicit]
        table [y error plus = Corr-Max, y error minus = Corr-Min]{\ytable};

    \legend{X-Coordinates, Y-Coordinates}
    \end{axis}
    \end{tikzpicture}
}
    \vspace*{-8pt}
    \caption{Average PPMC scores per each PT when masking 3 PTs at a time and for a model trained with $N$=3.}
    \vspace*{-10pt}
    \label{barchart:err_bars_N=3}
\end{figure}
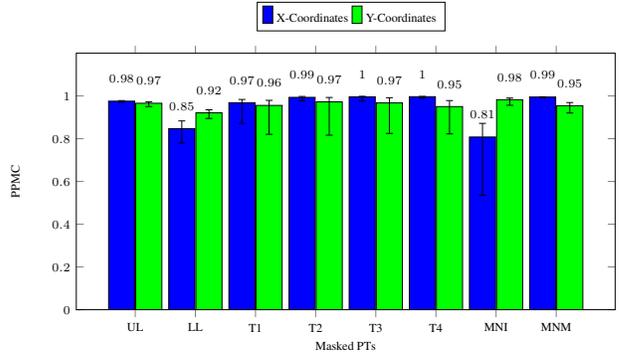

% \vspace{-10pt}
So far, we've shown the average PPMC score per level of masking during testing. This viewpoint, while useful for comparing models and hyper-parameter tuning, does not demonstrate enough of the consistency of the models' performance in retrieving different PTs. For this, we show the results from the same model from Figure \ref{barchart: overlap, N=3} in a more comprehensive way. Figure \ref{barchart:err_bars_N=3} shows the average PPMC score per PT when masking 3 PTs during training and with $N$=3. Each bar shows the average PPMC of each PT over all the combinations of 3 masked PTs that specific PT appears in. The error bars show the highest and lowest PPMC scores for each PT. The average PPMC score over all the cases in Figure \ref{barchart:err_bars_N=3} corresponds to the third bar in Figure {\ref{barchart: overlap, N=3}}.

\pgfplotstableread{
PT y Corr-Max Corr-Min
UL    0.975534    0.002781    0.005066
LL    0.847110    0.036034    0.067762
T1    0.968614    0.015019    0.097325
T2    0.993442    0.004182    0.016265
T3    0.995890    0.002572    0.020182
T4    0.996609    0.001217    0.007309
MNI    0.807900    0.063871    0.271959
MNM    0.994992    0.001353    0.003436
}{\xtable}
\pgfplotstableread{
PT y Corr-Max Corr-Min
UL    0.965809    0.007195    0.015223
LL    0.921009    0.014348    0.026235
T1    0.955690    0.023743    0.135063
T2    0.972533    0.020349    0.156005
T3    0.967496    0.024328    0.142664
T4    0.950224    0.028176    0.127662
MNI    0.982495    0.007671    0.026055
MNM    0.954489    0.014649    0.034214
}{\ytable}

It is clear from Figure \ref{barchart:err_bars_N=3} that the model does not handle some cases as well as others. In some cases the model is not able to retrieve missing PTs when closely related PTs are masked as well. For example, when UL and LL are masked together, the model often suffers to retrieve one or both of them. Similarly when MNI and MNM are both masked. This is also the case when masking three out of the 4 tongue PTs (T1, T2, T3 and T4). 

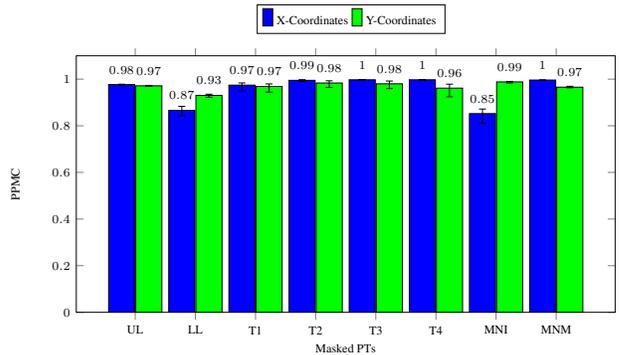
\begin{figure}[]
    \centering
\resizebox{0.95\columnwidth}{!}{%
    \begin{tikzpicture}[
      declare function={
        barW=4pt; % width of bars
        barShift=barW/2; % bar shift
      }
    ]
    \pgfplotsset{%
    height=0.38\textwidth,
    width=.7\textwidth
    }
    \begin{axis}[
        ybar = 0.6pt,
        ylabel=PPMC,
        xlabel = Masked PTs,
        ymin =0, ymax=1.1,
        enlarge x limits = 0.14,
        bar width=15pt,
        legend style={at={(0.51,1.2)},
        anchor=north,legend columns=0},
        symbolic x coords={UL, LL, T1, T2, T3, T4, MNI, MNM},nodes near coords,
        nodes near coords,style={font=\scriptsize, /pgf/number format/.cd,precision=2},      
        every node near coord/.append style={yshift = 0.1cm},
        xtick=data,         
        ytick={0,0.2,0.4,0.6,0.8,1},
        legend image code/.code={
        \draw [#1] (0cm,-0.1cm) rectangle (0.2cm,0.25cm); }]
    
        \addplot[ybar,fill=blue] plot [error bars/.cd, y dir= both, y explicit]
        table [y error plus = Corr-Max, y error minus = Corr-Min]{\xtable};
        \addplot[ybar,fill=green] plot [error bars/.cd, y dir= both, y explicit]
        table [y error plus = Corr-Max, y error minus = Corr-Min]{\ytable};

    \legend{X-Coordinates, Y-Coordinates}
    \end{axis}
    \end{tikzpicture}
}
    \vspace*{-8pt}
    \caption{Average PPMC scores per each PT when masking 3 PTs at a time and for a model trained with N=3. The test cases here do not include masking UL and LL, or MNI and MNM together, or three of the 4 tongue PTs.}
    \vspace*{-15pt}
    \label{barchart:err_bars_N=3-limiyed}
\end{figure}

Fortunately, these cases are not common in the subset of 41 speakers used in this study. Eliminating recordings with related PTs corrupted together reduces the duration of the reconstructed dataset by 0.12 hours, meaning that we can still increase the size of usable data in this subset from 7.2 hours to 10.48 hours. Disregarding these cases ensures that the model can reconstruct up to 3 concurrently corrupted PTs. Figure \ref{barchart:err_bars_N=3-limiyed} shows that limiting the test cases results in a more consistent performance with a worst-case PPMC of about 0.8 for speaker JW11. The worst-case PPMC score over the entire dataset is above 0.7, and the average PPMC score is above 0.9.

\pgfplotstableread{
PT y Corr-Max Corr-Min
UL    0.976465    0.001850    0.002080
LL    0.865704    0.017440    0.022931
T1    0.974122    0.009510    0.024869
T2    0.994811    0.002814    0.004184
T3    0.997105    0.001357    0.001643
T4    0.997073    0.000752    0.002072
MNI    0.852048    0.019722    0.042683
MNM    0.996022    0.000324    0.000546
}{\xtable}
\pgfplotstableread{
PT y Corr-Max Corr-Min
UL    0.971238    0.001766    0.001806
LL    0.930287    0.005070    0.007819
T1    0.968586    0.010847    0.024539
T2    0.983300    0.009582    0.019135
T3    0.979842    0.011982    0.020493
T4    0.961293    0.017107    0.036930
MNI    0.987987    0.002179    0.004306
MNM    0.965486    0.003652    0.002459
}{\ytable}
\section{Conclusion And Future Work}
\label{sec:conclusion}
% \vspace{-10pt}
We have shown that a Masked Autoencoder model can closely reconstruct mistracked  PTs in the XRMB dataset. We have expanded the usable portion from 7.2 hours to 10.48 hours, a 45\% increase in duration.

One limitation of our approach was its inability to confidently reconstruct concurrently mistracked related PTs. Future work should explore different architectures to further exploit interdependencies between different articulators. Furthermore, our data driven and speaker dependent approach meant we were unable to reconstruct mistracking in scarce datasets. A speaker independent approach should be invistigated in the future, along with data augmentation techniques and utilizing corrupted recordings in the training data.
\bibliographystyle{IEEEbib}
\bibliography{mybib}

\end{document}